# TOWARDS A GLOBAL SCIENTIFIC BRAIN:
## INDICATORS OF RESEARCHER MOBILITY USING CO-AFFILIATION DATA

**Cassidy R. Sugimoto[1]\*, Nicolás Robinson-Garcia[2], and Rodrigo Costas[3]**

[1] School of Informatics and Computing
Indiana University Bloomington, USA
sugimoto@indiana.edu
\*corresponding author

[2]INGENIO (CSIC-UPV)
Universitat Politècnica de Valéncia, Spain
elrobin@ingenio.upv.es

[3]Centre for Science and Technology Studies (CWTS)
Leiden University, The Netherlands
rcostas@cwts.leidenuniv.nl

*"Mobility—and in particular international mobility—of skilled human resources plays an important role in innovation. It contributes to the creation and diffusion of knowledge, particularly tacit knowledge, which is more effectively shared within a common social and geographical context."* --OECD (2010)

**Motivation**. Mobility of highly skilled workers has increased substantially since the 1990s. This increased mobility has a major impact on science and technology across the world (OECD, 2001; OECD, 2008) and is considered a central element in contemporary globalization (OECD, 2008; Docquier & Rapoport, 2012). The mobility of human capital is particularly important not only for the explicit sharing of resources, but also the exchange of tacit knowledge (Akerblom, 1996; OECD, 2008). Considerable attention, therefore, has been paid to the asymmetric flow of scholars between sending and receiving countries, with the latter tending to be economically wealthier (OECD, 2001; Docquier & Rapoport, 2012). The disparities caused by this asymmetry have been studied as brain drains and gains, in which receiving countries obtain a greater share of scientific capital to the detriment of the sending countries. Of particular concern has been the concentration of scholars in countries such as the United States. To account for these potential losses, many countries have created incentives for highly skilled scholars to stay in the country or to return after stays abroad (e.g., the Federation Fellowships in Australia) (Laudel, 2005). For example, in 2001, Science Foundation Ireland gave 10 scientific elite $67 million and planned another $530 million to try to retain Irish scholars and attract additional scientific talent (Pickrell, 2001).

However, contemporary scholars tend to argue that scientific mobility is not necessarily a zero-sum game (OECD, 2008); rather, that there are benefits that accrue to both sending and receiving countries (Docquier & Rapoport, 2012; OECD, 2001; OECD, 2008; Jonkers & Cruz-Castro, 2013; Jonkers & Tijssen, 2008), with temporary migration having particularly positive effects. This has stimulated several policies and incentive programs which urge scholars to engage in temporary work abroad (e.g., the Fulbright program in the US and the Ramón y Cajal in Spain). However, even those who remain abroad create peripheral benefits to the sending country (OECD, 2008; Jonkers & Cruz-Castro, 2013; Velema, 2012). As noted by the OECD (2008): "A stock of skilled HRST abroad can act as a conduit for flows of knowledge and information back to the home country, and social and other links increase the probability that knowledge will continue to flow back even after individuals move back or move away". However, these networks have been infrequently studied, due to the nature of mobility data.

***The construction of internationally comparable mobility indicators for the scientific workforce is a persistent policy need*** (Auriol, 2006; Moguerou et al., 2006; OECD, 2008). As summarized by an OECD (2008) report on the mobility of the highly skilled:

> "While recent years have seen major efforts to improve data on international stocks and flows of the highly skilled, difficulties relating to international comparability, to differing and/or insufficient disaggregation and to timeliness remain. Further work is needed if countries are to better understand patterns and changes in stocks and flows of scientists, engineers, and researchers and the broader category of the highly skilled…Quantitative evidence on the impact of mobility patterns is not readily available. Many variables and factors influence science and technology outcomes and are hard to disentangle" (OECD, 2008).

**Previous approaches to studying mobility.** In a comprehensive review of mobility indicators, Åkerblom (2002) concluded that "the possibilities for constructing indicators on international mobility do not look very promising" based on issues of data sparsity, consistency, and interoperability. Data on mobility largely comes from census data, registration data, labor force surveys (e.g., CLFS), longitudinal panels (e.g., NCSES, EuroSTAT, ECHP), individual and organization surveys, and specific case studies—none of which have been seen to be sufficient in providing comprehensive and contemporary analysis of scientific migration for policy purposes (Åkerblom, 1996; OECD, 2001; Moguerou et al., 2006; Cañibano & Bozeman, 2009).

Conceptualizations have also hindered indicator development: mobility has been largely measured in terms of *stock*, rather than *flow* (OECD, 2010)—measuring binary distinctions between domestic and international participants in the scientific workforce at a particular time and within a particular region. These data fail to account for the migratory pattern of scholars and how, in their movements, they create networks of relationships among institutions and countries over time. Longitudinal panels show more interaction, but are subject to sampling and response bias (Åkerblom, 1996). For all of these, there is a delay in obtaining data and aggregating the data (Moguerou et al., 2006), which results in highly dated reporting for mobility indicators (Docquier & Rapoport, 2012). Furthermore, as aggregate data, they fail to tell data about individual trajectories, therefore suppressing data about sociodemographic characteristics or metrics of scientific capital.

Data on the "highly skilled" are non-interoperable and difficult to directly compare, given that data are collected in idiosyncratic ways. The Canberra Manual provides a very broad definition of highly skilled workers (human resources in science and technology (HRST) in OECD parlance (OECD, 2008)). Others have used the ISCO (International Standard Classification of Occupations) classification, but this has been variably adopted. Furthermore, there is ambiguity within these classification systems on what constitutes a high skilled or "highly qualified" individual (Akerblom, 1996). These ambiguities in classification impeded the development of globally comparable indicators. Furthermore, the scientific workforce constitutes a subset of so-called "highly-skilled" workers. However, most large-scale data collection activity aggregates these individuals. Many studies, therefore, report on migration behaviors of the highly-skilled, without isolating the scientific workforce (e.g., Suzuki & Suzuki, 2016). Such studies may over-represent certain disciplines and sectors of work while suppressing the activities or researchers in disciplines, such as the social sciences and humanities.

**Bibliometric approach to studying mobility**. As an alternative, we propose using affiliation and co-affiliation data from scientific publications to infer mobility patterns. This deviates from traditional collaboration studies in that it examines *individuals associated with multiple institutions* simultaneously and over time, rather than the association of institutions through collaboration of people with distinct institutional affiliations (e.g., Gazni, Sugimoto, & Didegah, 2012). Using Thomson Reuters' Web of Science (WoS) database, this can be calculated at a global scale and can be contextualized with inferred data about age (Costas et al, 2015) and gender (Larivière et al, 2013) and analyzed at various levels of aggregation (i.e., country, region, institution, city). Given that publications are associated with date information, we can conduct diachronic network analyses to identify the trade of scholars not only between locations, but among all locations over time. Publication data provide the added advantage that we can examine the impact of mobility, by measuring citations before, after, and during periods of transition. Furthermore, bibliometric data can be analyzed at least quarterly, which addresses the problem of the delays in obtaining statistics on R&D personnel that has been repeatedly noted in the literature (e.g., Moguerou et al., 2006).

Several unobtrusive methods have been used to derive mobility data for scientists: e.g., encyclopedias, biographical information, and curriculum vitae (Laudel, 2003). The use of these data, however, is inherently limited in scope, coverage, and accuracy and requires a substantial amount of cleaning, reducing the utility of the datasets for large-scale analyses (see, e.g., Dietz et al., 2000). Laudel (2003, 2005) was one of the first to argue for the use of bibliometrics to construct global indicators on scientific mobility. Laudel (2005) employed PubMed data, which provides first-author affiliations since 1980. She augmented her analysis with data on affiliation of doctoral degree. However, she restricted her analysis to a certain classification of elite scholars: that is, those who published at least three papers in *Science* and *Nature* between 1980 and 2002. This is a restrictive dataset, both in terms of disciplinary coverage as well as country coverage. Taking only first-author data suppresses the contributions of the scientists who labor in large collaborative teams. Looking at those who have published three times in elite journals further reduces the pool of eligible scientists. This is a persistent problem in the literature on mobility, with many studies focusing on the elite and super elite—for example, Nobel Prize winners (Zuckerman, 1995; Hunter, Oswald, & Charloton, 2009). Zuckerman (1995) defined the elite as those who "have made a difference to the advancement of scientific knowledge". However, there is a great deal of subjectivity in what constitutes "a difference". As Laudel noted, these definitions neglect "the conditions of collective production" (Laudel, 2005, p. 381). Here we take a more prosopographical approach: studying all those who contribute to science, regardless of their status in the research system.

Our work builds upon the research of Moed and Halevi (2014), who examined migration balances between a select group of developing and developed countries, using Scopus data. This work was methodologically useful in that it both discussed and examined the difficulties of author-name disambiguation for bibliometric data, including complexities of homonyms and synonyms in the database. In their validation study of 100 randomly selected Chemistry authors, they conclude that database errors contribute to relatively little change in the results. Our work also draws upon a study Markova and colleagues (2016), who proposed the notion of "synchronous mobility" to describe those holding two simultaneous posts and provided a proof-of-concept analysis using affiliation data of Russian scientists from Web of Science. Our proposed method takes both synchronous (i.e., co-affiliation) and diachronous (i.e., movement from one entity to another) mobility into account.

**Methods**. We use a large-scale author disambiguation algorithm applied to the in-house CWTS version of the WoS database (Caron & van Eck, 2014). This algorithm clusters publications that belong to a single author based on a set of similar publication patterns (co-author network, address data, etc.). This allows us to track the whole publication output of each author. We consider all researchers with at least two publications whose first paper dates from 2003 onwards. We track their publication record for the 2003-2015 period.

Several variables were collected to aid in the analysis, including:
- Number of publications;
- Years of the first and last publications of the authors;
- Countries with which the authors has been affiliated;
- Publication years associated with each affiliation;
- Origin affiliation: i.e., affiliation at time of first publication; and,
- Total, mean, and normalized citation scores (including proportion of publication in the top 10% of most cited publications within their field).

We identify three types of mobility events for each given author-publication and year-entity combination: 1) no mobility, 2) mobility (based on production in another country different from the preceding time period), and 3) multiple affiliation (affiliated with multiple countries simultaneously). We also calculate "return"—that is, when mobility returns to the origin affiliation.

Based on this dataset, we identify for each researcher her two most common affiliations. Such restrictive assignment of affiliations allows us to rely only in trusted links between affiliations. For each affiliation the organization name, city, and country were identified, allowing us to create co-occurrence matrices based on the number of researchers who share common organization/cities/countries. Then we create co-affiliation networks at three levels of aggregation: country, city, and organization. For the first case study we employ the complete WoS database. Mobility networks at the city level are performed at the national level for the USA, Spain, and The Netherlands, as a proof-of-concept exercise. Organization networks are shown only for Spain and The Netherlands. In this case, due to the messiness of the institutional names data (Robinson-Garcia & Calero-Medina, 2014), we performed data cleaning in order to remove duplicities. Finally, we used the VOSViewer software for visualizing the networks (van Eck & Waltman, 2010). Centrality measures are reported in tabular form to assist in interpretation.

For each country, we also calculated the total number of scholars associated with the country and calculated the proportion of scholars that were "sent" or "received" (by analyzing directionality of mobility events). The total number of scholars provides a capacity indicator for the country which can be used to calculate both expected shares and normalized shares of immigrating and emigrating scholars. We plot these normalized shares both for sending and receiving countries to demonstrate the flow of scholars within Europe.

**Proof of concept.** A descriptive visualization of the co-affiliation data confirms our hypothesis that countries will group in regional and cultural clusters (Figure 1). Node sizes represent the number of researchers affiliated to a given country/city/organization. Edges represent researchers who have reported in their publications affiliation to two given nodes. In this case we show non-normalized networks as we work with relational, rather than positional, networks. The United States occupies a central position in the network, closely surrounded by Canada, England, Scotland, and Germany. Spain and Brazil, although they are of equal size to Canada, Germany, and England are fairly separated in the network forming a cluster with Latin-American countries. China is large, but also peripheral and predominately connected to other Asian countries. It connects at the top left of the graph with Asian and north African countries of smaller presence, which in turn connect with Turkey, Iran, and other Muslim countries. Here Turkey connects with Russia. Finally, at the bottom of the network, we find the African countries clearly separated from North Africa, with South Africa connecting them with the rest of the network.

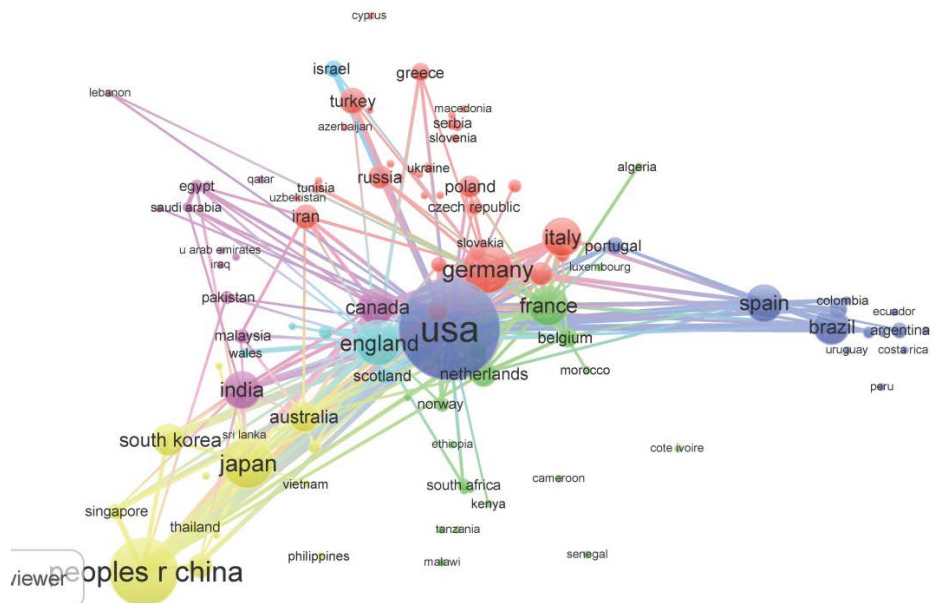

Figure 1. Co-affiliation network (non-normalized) on the country level (2003-2014)

Calculating centrality measures at the world and regional level can provide further insights on those countries serving as conduits for mobility (Table 1). At the world level, we see that the United Kingdom occupies the place of highest centrality, suggesting this is a bridge between countries in terms of scientific mobility. European countries dominate this list, representing more than half of the top 15 bridging countries, in terms of betweenness centrality. However, Australia also occupies a central position, with Oceania serving as a bridge between Asia and Europe.

Table 1. Centrality measures for the top 15 countries, based on betweenness centrality measures (2008-2015)

| Country | # of researchers | Closeness | Betweenness |
|---|---|---|---|
| **UNITED KINGDOM** | 186054 | 0.811538 | 4090.422 |
| **USA** | 730612 | 0.871901 | 2845.715 |
| **AUSTRALIA** | 73165 | 0.727586 | 1723.924 |
| **FRANCE** | 118623 | 0.770073 | 1528.891 |
| **ITALY** | 100091 | 0.705686 | 789.6576 |
| **SPAIN** | 105871 | 0.674121 | 775.1448 |
| **GERMANY** | 177979 | 0.730104 | 724.8132 |
| **SWITZERLAND** | 38669 | 0.703333 | 646.2715 |
| **CANADA** | 96468 | 0.722603 | 567.864 |
| **NETHERLANDS** | 45939 | 0.700997 | 484.6474 |
| **INDIA** | 110536 | 0.680645 | 465.3142 |
| **PEOPLES R CHINA** | 409957 | 0.661442 | 437.7965 |
| **JAPAN** | 182467 | 0.671975 | 410.6764 |
| **BELGIUM** | 27233 | 0.694079 | 391.7732 |
| **SENEGAL** | 737 | 0.548052 | 256.1233 |

Centrality measures can also be calculated at the level of a given region. For example, Table 2 demonstrates the importance of South Africa as a bridging country within Africa.

Table 2. Centrality measures for the 10 most central African countries, based on betweenness (2008-2015)

| Country | # researchers | Closeness | Betweenness |
|---|---|---|---|
| **SOUTH AFRICA** | 13416 | 0.714286 | 321.8305 |
| **SENEGAL** | 737 | 0.588235 | 146.9085 |
| **NIGERIA** | 5647 | 0.632911 | 115.5662 |
| **CONGO** | 410 | 0.581395 | 87.39322 |
| **GHANA** | 1425 | 0.588235 | 83.71863 |
| **CAMEROON** | 1476 | 0.549451 | 81.68561 |
| **KENYA** | 2820 | 0.632911 | 77.61292 |
| **UGANDA** | 1674 | 0.625 | 75.93211 |
| **SUDAN** | 685 | 0.480769 | 49.82798 |
| **ETHIOPIA** | 1702 | 0.574713 | 44.20708 |

One can also identify coherent results at a finer level of analysis. For example, three countries are presented here, with data at the level of city (Figure 2).

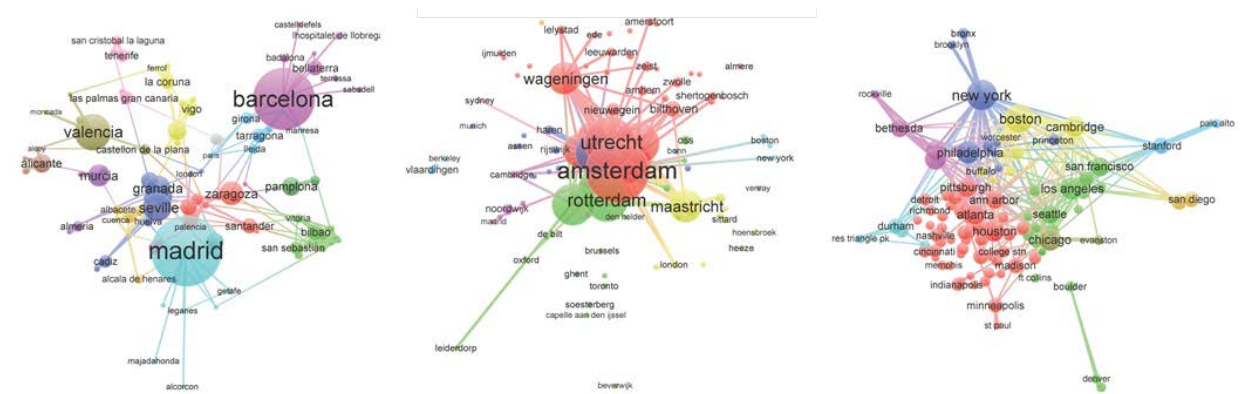

Figure 2. Co-affiliation networks (non-normalized) at the city level for Spain, The Netherlands, and the United States (2003-2014)

Noticeable here are the strong regional components for co-affiliations in the United States and Spain (e.g., Madrid, Catalan, Andalusian, Valencian clusters for the Spain map, and East/West distinctions in the US). However, the Dutch map does not contain this regional component, showing a close network of the main university cities of the Randstad (e.g. Amsterdam, Utrecht, Rotterdam, Leiden, etc.), thus exhibiting a lower regional component compared to the other two countries. Furthermore research is needed to understand what contextual factors within a country (e.g., size, R&D expenditure, etc.) contribute to these clusterings.

Of course, several institutions are often in a single city. Therefore, to examine institutional exchange, one can present these data at the level of organization (Figure 3). Following we show the top organizations with the highest number of researchers in The Netherlands and Spain. For both countries, it is important to note the important role of hospitals which are orbiting in all cases around the main universities of the cities in which they are located, for example the Erasmus University of Rotterdam and the Erasmus Medical Center, or the University of Barcelona and the Hospital Clinic from Barcelona. The Dutch network (left) shows a mix of geographical (e.g. University of Amsterdam, Free University of Amsterdam and Leiden University close in the map) and thematic (e.g. the Technical universities, Delft, Eindhoven and Twente clustered together) patterns in the co-affiliation relationships. In the Spanish case (right), is noticeable the central role of the Spanish Research Council (CSIC) in the network. We observe also a strong regional pattern with institutions from Madrid at the top-left of the network and Catalan institutions at the top-right. Basque universities cluster with institutions from Navarre, while Andalusian institutions are tightly clustered beneath the CSIC node, and there is also a cluster (dark blue) with the universities from the north-west of Spain (Galicia, Asturias, Cantabria).

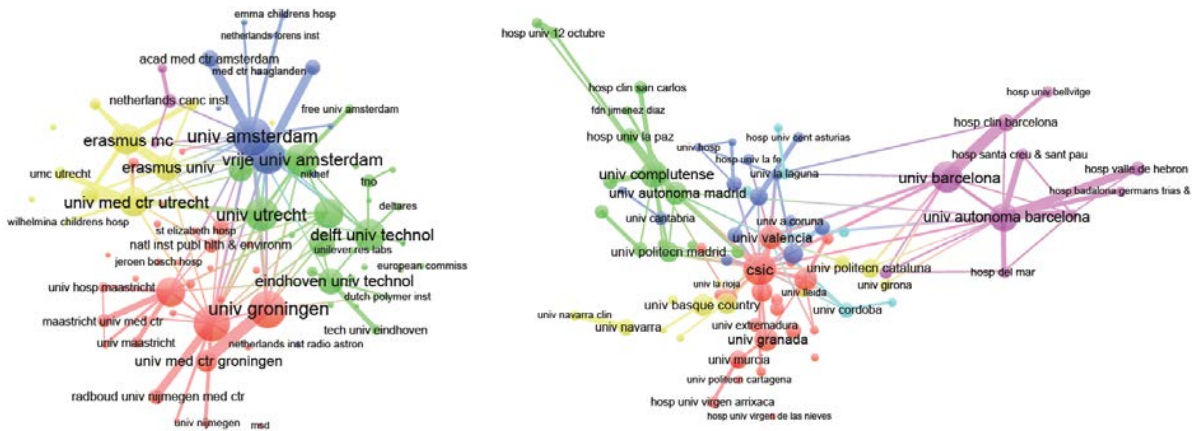

Figure 3. Co-affiliation networks (non-normalized) at the institution level for The Netherlands and Spain (2003-2014)

Co-affiliation provides an indicator of bridging units. However, it is also important to recognize those countries which tend to be importing and exporting scholars. To this end, we calculate a normalized share of imported and exported scholars by country. As a proof-of-concept, data for Europe is provided. Figure 4 provides a heatmap of the normalized share, where the darker regions indicate those who are sending more than their expected share of scholars to other regions. As shown, Eastern Europe features prominently as a "sending" region.

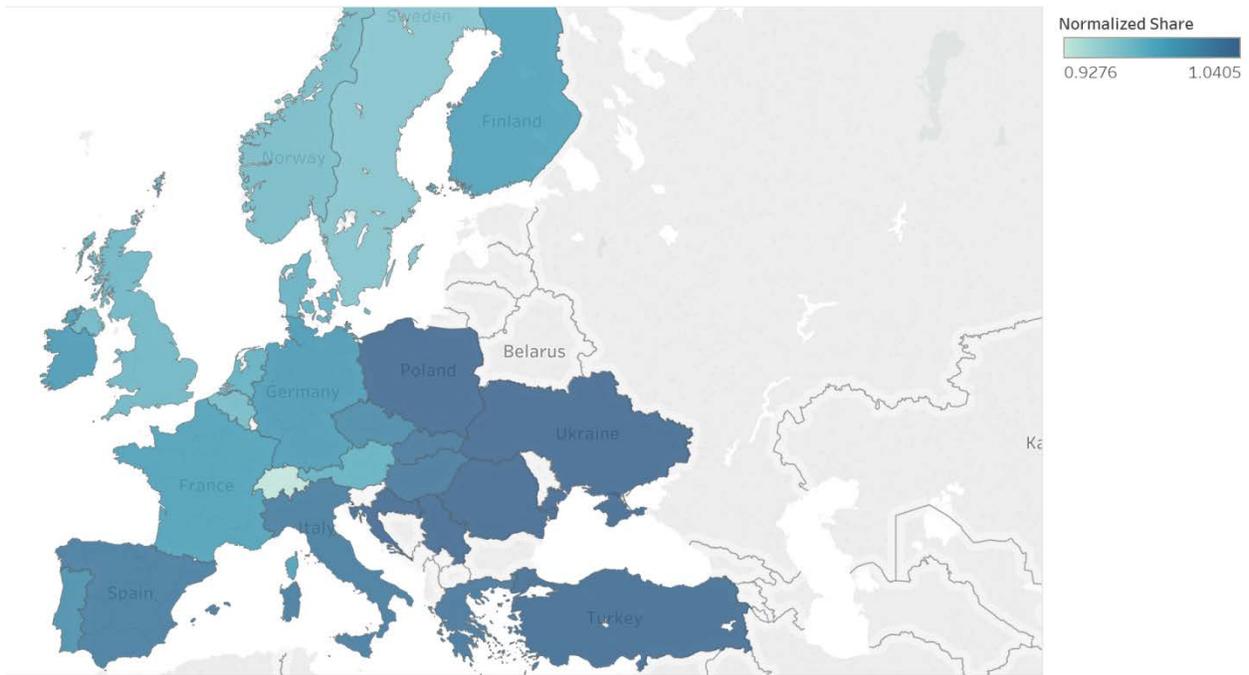

Figure 4. Normalized share in terms of sending authors

Figure 5 provides the normalized share in terms of receiving authors—that is, this indicates the countries to which scholars emigrate. As shown, Scandinavian countries and Switzerland receive more researchers than would be expected, given the size of their respective researching populations. Several explanations can be made for this—for example, one might argue that the resources in CERN are a major draw for physicists. The value-add of using bibliometric data as a mobility indicator is that such hypotheses can be easily tested, by examining the field and even topics of the papers which are mobile or co-affiliated.

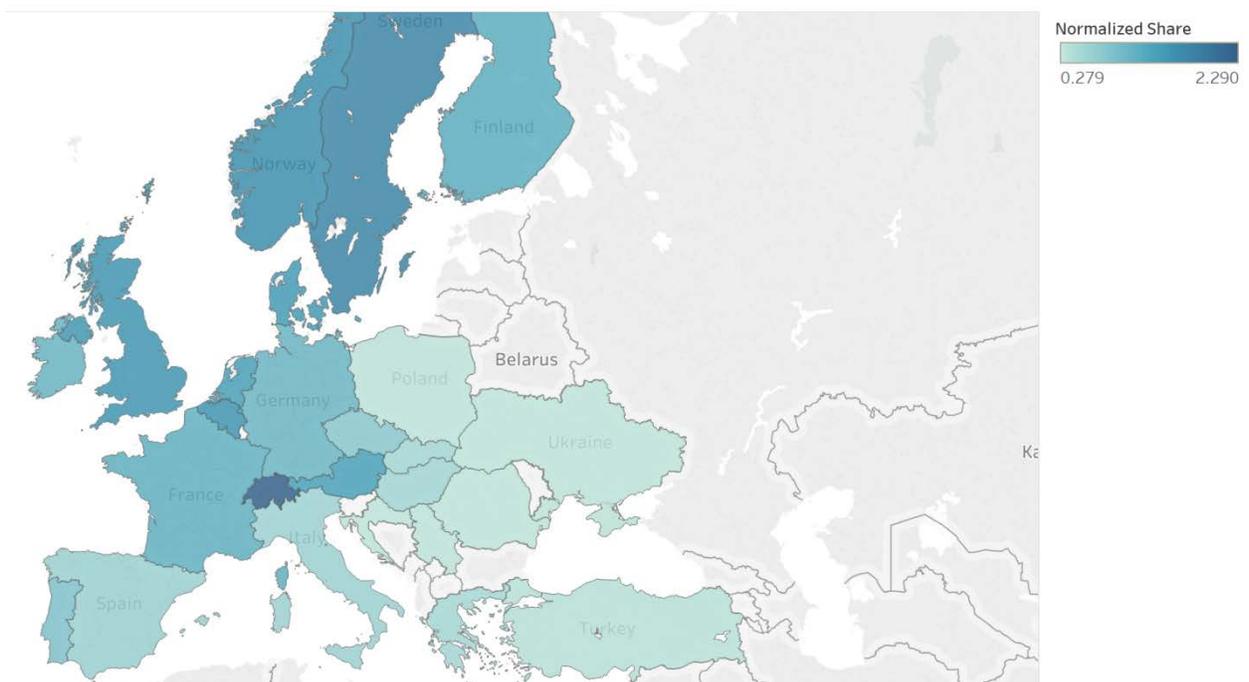

Figure 5. Normalized share in terms of receiving authors

It is also of great policy interest to study the effects, both in terms of productivity and impact, of researchers who are mobile. Bibliometric data allows for this analysis to the level of each individual publication. To demonstrate, we calculate the mean normalized citation score (MNCS) for papers identified as "mobility", "mobility and multiple affiliation", and "multiple affiliation" (Figure 6). In all cases, such papers received a higher MNCS than those papers not associated with mobility. This indicator could be applied at any level of analysis (i.e., city, institution, country, field).

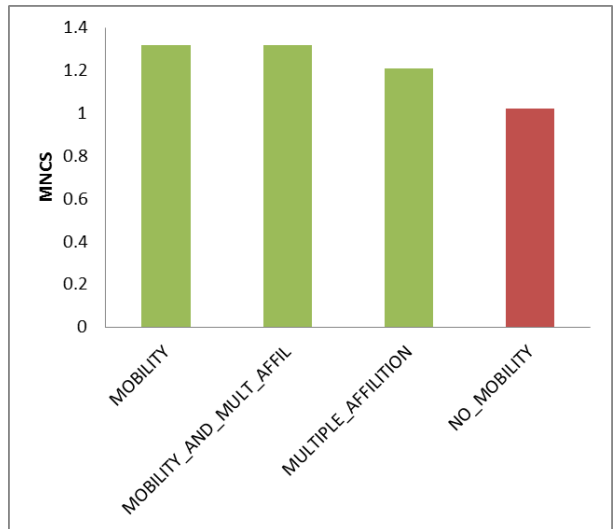

Figure 6. Mean normalized citation count by type of mobility event

**Limitations to the approach.** There are several strengths to this method, which overcome the weaknesses of previous methods of establishing internationally comparable indicators on the mobility of scientists. However, bibliometric data has inherent limitations that must be taken into account. Some of the well-known caveats of bibliometrics apply here: that is, the issues in coverage for language, country, and discipline that persist in Web of Science and other large-scale indexes (Cronin & Sugimoto, 2015).

Another piece of information that is not available from citation indexes alone is the place of doctoral education, an important element as the prestige of graduate school has been shown to influence first appointments (Debackere & Rappa, 1995). Previous studies have used place of first publication as a proxy for institution of doctoral education (Moed & Halevi, 2014; Laudel, 2003). Matching Web of Science data with other data, such as ProQuest's Dissertations and Theses database, could alleviate some of these problems. However, this database is highly skewed toward North America, which would diminish the global coverage of the indicator. Therefore, more validation studies should be conducted to estimate the accuracy of inferring doctoral education data from place of first publication.

Finally, it should be remembered that this approach is primarily descriptive, rather than explanatory. That is, we provide a globally standard and timely way of demonstrating knowledge flows of human capital, but we cannot understand all the factors that may motivate this mobility. Therefore, this data should not supplant, but rather supplement, the current collection of motivational factors done with smaller-scale investigations. Together, these multiple approaches provide a more comprehensive policy picture.

**Implications**. It has been twenty years since the first OECD Blue Skies conference in which Akerblom (1996) bemoaned the paucity of internationally comparable data sets on the mobility of highly skilled workers. Ten years later, at the second OECD Blue Skies conference, these concerns were reiterated (Auriol, 2006). Our proposed technique circumvents many of the previous issues by adopting a simple proxy—that is, authorship—for a research and using a uniform measurement for mobility—that is, affiliation on papers, which can be analyzed both contemporaneously and historically.

This work has several data and infrastructure implications as well as policy implications. The internationality of institutions is being widely used in rankings, but tends to focus on collaboration over mobility—that is, the emphasis is on the work products, rather than on the individuals involved in knowledge transfer. Constructing of more nuanced indicators for the 21$^{st}$ century must focus more on human capital and humans as actors in knowledge transfer. Previous analyses have relied on complex integration of data from several sources; making indicator development cumbersome and dependent upon the continually systematic collection of data from heterogeneous sources.

Our knowledge of stocks of human capital is fairly well-established through several national surveys and registries, though the quality and consistency of data varies by country. What is less well understood, however, are the flows of human capital across countries, cities, and institutions. Furthermore, it is not understood whether benefits to mobility are universal—notions such as brain drain and brain gain suggest unidirectionality of benefit; but the relative cost/benefit by country is unknown. Furthermore, the scientific and policy communities are promoting globalization, but lack clear indicators on mobility. ***What is needed, therefore, are global data sources, validated indicators, and contextualized interpretations that allow us a more refined understanding of the effects of mobility on science***. The approach proposed in this study opens the possibility for simple and internationally comparable indicators to be devised, although limitations of bibliographic data for migration and mobility analysis must also be regarded (Moed et al, 2013).

Our approach includes students in the population of migrating scientists. This is important due to the role that these individuals may play in both the production of science (Larivière, 2012) as well as their roles in linking agents between countries (Akerblom, 1996; OECD, 2008). In a study of exceptional contributions to science and engineering in the United States, Levin and Stephan (1999, 2001) found that the contributions were disproportionately made by researchers who were both foreign born and foreign educated. This story can be read in several ways: on the one hand, it can be interpreted that the US provided these scholars the resources necessary to flourish. But it can also be demonstrated that the US benefitted significantly from the educational investments of other countries. It is clear that what are necessary are indicators, at the global scale, of the exchange of scientific capital.

We must also move from a notion of mobility to *mobilities* (Laudel, 2005), considering mobility across sectors (e.g., Janger & Nowotny, 2016; Sandgren & Perez, 2006), within a given country (e.g., from rural to urban settings) (Suzuki & Suzuki, 2016), and internationally (Akberlom, 1996). In only some of these cases is mobility also migration (Laudel, 2003, 2005). Studies of mobility have focused on specific countries (e.g., Suzuki & Suzuki, 2016; Kim, 2006; Jonkers & Cruz-Castro, 2013; Velema, 2012; Jonkers & Tijssen, 2008) or those within a specific subspecialty (Laudel, 2003). Such studies provide important policy implications at the national or disciplinary level, but fail to examine the global significance of these migratory patterns.

**Imagining indicators**. This paper has advocated for the use of bibliometric data to construct indicators for scientific mobility. A proof-of-concept demonstration was provided; however, there are several other indicators that can be constructed on the basis of these data that have significant policy implications. There is a tremendous amount of information embedded in citation indexes which can be overlaid on mobility patterns. For example, it is hypothesized that there exists a strong relationship between the capital (social and scientific) possessed by a researchers and their propensity and ability to be mobile (Jonkers & Tijssen, 2008). Bibliometric data allows for the analyses of various correlative variables with mobility: both proxies for scientific capital (e.g., productivity and citation impact) as well as social scientific capital (e.g., prestige rankings of institutions or journals with which the author is affiliated). Small-scale studies have confirmed relationships among several of these variables: e.g., that productivity of a scholar affects the "destination prestige" of the receiving institution (Allison & Long, 1987) and potential for mobility (Van Heeringen & Dijkwel, 1987; Hunter, Oswald, & Charlton, 2009). Studies have also demonstrated the distinctions between collaboration and mobility networks (Furukawa, Shirakawa, & Okuwada, 2011). Bibliometric data could further elucidate the relationship among productivity, collaboration, impact, and mobility on a global scale.

Research has suggested that mobility of women may differ from that of men (Rosenfeld & Jones, 1987; Shauman & Xie, 1996; Docquier & Rapoport, 2012), though the extent and direction of these differences has found conflicting results due to non-equivalent samples. However, using name-gender algorithms, such as those established by Lariviere et al. (2013) for Web of Science data, one can easily overlay gender on these mobility patterns. Other sociodemographic characteristics, such as age (Costas et al., 2015), can also be inferred from the data and incorporated into these indicators.

There are several motivations for mobility, including better scientific infrastructure, scientific freedom, and the access to other researchers (OECD, 2008; Cañibano, Otamendi & Andújar, 2008). Therefore, indicators of mobility must not only analyze trade in stocks of highly skilled personnel, but the size of the population and potential pool of highly skilled workers who can emigrate as well as the capacity of the country to accept scientifically skilled workers (Hunter, Oswald, & Charlton, 2009; Docquier & Rapoport, 2012). As Laudel (2005) noted, "it is necessary to have elites to generate elites, it is also necessary to have attractive working conditions for potential elites." There are, simply put, differences in a country's ability to incentivize their local-born researchers to return home (Jonkers & Tijssen, 2008). Calculations of the "critical mass of native scientists" (Ioannidis, 2004) can be performed and modeled given bibliometric data. Furthermore, it has been argued that the global stock of knowledge increases when scientists move to areas where they have access to resources and expertise which propel their own productivity and innovation (OECD, 2008). Large-scale and globally comparable indicators make it possible to begin to test these hypotheses systematically and diachronically.

## ACKNOWLEDGEMENTS
The authors would like to thank Dakota Murray for his help in visualizing the sending and receiving countries.

## CITED REFERENCES
Åkerblom, M. (1996). Constructing internationally comparable indicators on the mobility of highly qualified workers. STI Review, 27. Retrieved from: http://www.oecd.org/sti/37124998.pdf?utm_source=&utm_medium=newsletter&utm_content=618456&utm_campaign=STI-News-December-2015

Allison, P.D., & Long, S.J. (1987). Interuniversity mobility of academic scientists. American Sociological Review, 52 (5), 643-652.

Auriol, L. (2006). International mobility of doctorate holders: first results and methodology advances. "Blue Skye II 2006": What indicators for science, technology and innovation policies in the 21st century? Retrieved from: http://www.oecd.org/sti/inno/37450315.pdf

Cañibano, C. & Bozeman, B. (2009). Curriculum vitae method in science policy and research evaluation: the state-of-the-art. *Research Evaluation*, 18(2), 86– 94.

Caron, E., & van Eck, N. J. (2014). Large scale author name disambiguation using rule-based scoring and clustering. In *19th International Conference on Science and Technology Indicators.'Context counts: Pathways to master big data and little data'* (pp. 79–86).

Costas, R., Nane, T., & Larivière, V. (2015). Is the Year of First Publication a Good Proxy of Scholars' Academic Age?. In *15th International Conference on Scientometrics & Informetrics*. Istanbul (Turkey) (pp. 988-998). http://www.issi2015.org/files/downloads/all-papers/0988.pdf

Cronin, B., & Sugimoto, C.R. (2015). Scholarly metrics under the microscope: from citation analysis to academic auditing. Medford, NJ: Information Today, Inc./ASIST.

Debackere, K. & Rappa, M.A. (1995). Scientists at major and minor universities: mobility along the prestige continuum. Research Policy, 24, 137-150.

Dietz, J.S., Chompalov, I., Bozeman, B., O'Neil Lane, E., & Park, J. (2000). Using the curriculum vita to study the career paths of scientists and engineers: An exploratory assessment. *Scientometrics*, *49*, 419-442.

Docquier, F., & Rapoport, H. (2012). Globalization, brain drain, and development. Journal of Economic Literature, 50(3), 681-730.

Furukawa, T., Shirakawa, N., Okuwada, K. (2011). Quantitative analysis of collaborative and mobility networks. *Scientometrics, 87*, 1–16.

Gazni, A., Sugimoto, C.R., & Didegah, F. (2012). Mapping world scientific collaboration: Authors, institutions, and countries. Journal of the American Society for Information Science & Technology, 63(3), 450-468.

Hunter, R.S., Oswald, A.J., & Charlton, B.G. (2009). The elite brain drain. The Economic Journal, 119(538), F231-F251.

Ioannidis ,J.P.A. (2004). 'Global estimates of high-level brain drain and deficit'. The Journal of the Federation of American Societies for Experimental Biology, 18(9), 936-939.

Janger, J., & Nowotny, K. (2016). Job choice in academia. Research Policy, 45, 1672-1683.

Jonkers, K. & Cruz-Castro, L. (2013). Research upon return: The effect of international mobility on scientific ties, production and impact. Research Policy, 42, 1366-1377.

Jonkers, K., Tijssen, R. (2008). Chinese researchers returning home: impacts of international mobility on research collaboration and scientific productivity. *Scientometrics*, *77*, 309–333.

Kim, K.-W. (2006). Developing indicators for the effective utilization of HRST: The case of South Korea. "Blue Skye II 2006": What indicators for science, technology and innovation policies in the 21st century? Retrieved from: http://www.oecd.org/sti/inno/37450333.pdf

Larivière, V. (2012). On the shoulders of students? The contribution of PhD students to the advancement of knowledge. Scientometrics, 90(2), 463-481.

Larivière, V., Gingras, Y., Cronin, B., & Sugimoto, C. R. (2013). Global gender disparities in science. *Nature*, 504, 4–6.

Laudel, G. (2003). Studying the brain drain: Can bibliometric methods help? Scientometrics, 57(2), 215-237.

Laudel, G. (2005). Migration currents among the scientific elite. Minerva, 43(4), 377-395.

Levin, S.G., & Stephan, P.E. (1999). Are the foreign born a source of strength for US science? Science, 285(5431), 1213-1214.

Markova, Y.V., Shmatko, N.A., & Katchanov, Y.L. (2016). Synchronous international scientific mobility in the space of affiliations: evidence from Russia. SpringerPlus, 5, 480.

Moed, H. & Halevi, G. (2014). A bibliometric approach to tracking international scientific migration. Scientometrics, 101(3), 1987-2001.

Moed, H. F., Aisati, M. M., & Plume, A. (2013). Studying scientific migration in Scopus. *Scientometrics*, 94, 929–942. doi:10.1007/s11192-012-0783-9

Moguerou, P., Da Costa, O., Paola di Pietrogiacomo, M., & Laget, P. (2006). Indicators on researchers' career and mobility in Europe: A modeling approach. "Blue Skye II 2006": What indicators for science, technology and innovation policies in the 21st century? Retrieved from: http://www.oecd.org/sti/inno/37450342.pdf

OECD. (2001). *International mobility of the highly skilled*. OECD Publications.

OECD. (2008). The Global Competition for Talent: Mobility of the highly skilled. OECD Publications.

OECD. (2010), *Measuring Innovation: A New Perspective*, OECD Publishing, Paris.

Pickrell, J. (2001). Fighting brain drain: Ireland gives its stars a big pot o' gold. Science, 293(5532), 1028-1029.

Robinson-Garcia, N. & Calero-Medina, C. What do rankings by fields rank? Exploring discrepancies between the organizational structure of universities and bibliometric classifications. *Scientometrics, 98*(3), 1955-1970.

Rosenfeld, R.A. & Jones, J.A. (1987). Patterns and effects of geographic mobility for academic women and men. Journal of Higher Education, 58(5), 493-515.

Sandgren, P., & Perez, E. (2006). Mobility of the higher skilled in the Swedish Innovation System—An indicators for knowledge flows. "Blue Skye II 2006": What indicators for science, technology and innovation policies in the 21st century? Retrieved from: http://www.oecd.org/sti/inno/37450351.pdf

Shauman, K.A., & Xie, Y. (1996). Geographic mobility of scientists: Sex differences and family constraints. *Demography*, *33*(4), 455-468.

Stephan, P.E., & Levin, S.G. (2001). Exceptional contributions to US science by the foreign-born and foreign-educated. *Population Research and Policy Review*, *20*(1/2), 59-79.

Suzuki, Y., & Suzuki, Y. (2016). Interprovincial migration and human capital formation in China. Asian Economics Journal, 30(2), 171-195.

Van Eck, N.J., & Waltman, L. (2010). Software survey: VOSviewer, a computer program for bibliometric mapping. Scientometrics, *84*(2), 523-538.

Van Heeringen, A., & Dijkwel, P.A. (1987). The relationships between age, mobility and scientific productivity. Part I. Effect of mobility on productivity. Scientometrics, 11(5), 267-280.

Velema, T.A., 2012. The contingent nature of brain grain and brain circulation: their foreign context and the impact of return scientists on the scientific community in their country of origin. Scientometrics 93, 893–913.

Zuckerman, H. (1995). Scientific elite: Nobel Laureates in the United States. Transaction Publishers.